\documentclass[manuscript]{aastex63} 

\usepackage{bm}

\received{??} 
\revised{??} 
\accepted{??} 

\submitjournal{ApJL}

\shorttitle{Weak Solar Radio Bursts Observed by Parker Solar Probe}

\shortauthors{Chen et al.}

\begin{document}

\title{Weak Solar Radio Bursts from the Solar Wind Acceleration Region Observed by Parker Solar Probe and Its Probable Emission Mechanism}

\correspondingauthor{Ling Chen}
\email{clvslc214@pmo.ac.cn}

\author[0000-0001-8058-2765]{Ling Chen}
\affiliation{Key Laboratory of Planetary Sciences, Purple Mountain Observatory, 
Chinese Academy of Sciences, Nanjing 210023, People’s Republic of China}
\affiliation{State Key Laboratory of Space Weather, Chinese Academy of Sciences, Beijing 100190, People’s Republic of China}
\affiliation{CAS Center for Excellence in Comparative Planetology, Hefei 230026, People’s Republic of China}

\author[0000-0003-0888-8879]{Bing Ma}
\affiliation{Key Laboratory of Planetary Sciences, Purple Mountain Observatory, 
Chinese Academy of Sciences, Nanjing 210023, People’s Republic of China}

\author[0000-0003-2418-5508]{Dejin Wu}
\affiliation{Key Laboratory of Planetary Sciences, Purple Mountain Observatory, 
Chinese Academy of Sciences, Nanjing 210023, People’s Republic of China}
\affiliation{CAS Center for Excellence in Comparative Planetology, Hefei 230026, People’s Republic of China}

\author{Xiaowei Zhou}
\affiliation{Key Laboratory of Planetary Sciences, Purple Mountain Observatory, 
Chinese Academy of Sciences, Nanjing 210023, People’s Republic of China}

\author{Marc Pulupa}
\affiliation{Space Sciences Laboratory, University of California, Berkeley, CA 94720-7450, USA}

\author[0000-0001-6855-5799]{PeiJin Zhang}
\affiliation{Department of Physics, University of Helsinki, PO Box 64, 00014 Helsinki, Finland}
\affiliation{ASTRON, The Netherlands Institute for Radio Astronomy, Oude Hoogeveensedijk 4, 7991 PD Dwingeloo, The Netherlands}

\author{Pietro Zucca}
\affiliation{ASTRON, The Netherlands Institute for Radio Astronomy, Oude Hoogeveensedijk 4, 7991 PD Dwingeloo, The Netherlands}

\author[0000-0002-1989-3596]{Stuart D. Bale}
\affiliation{Space Sciences Laboratory, University of California, Berkeley, CA 94720-7450, USA}
\affiliation{Physics Department, University of California, Berkeley, CA 94720-7300, USA}
\affiliation{The Blackett Laboratory, Imperial College London, London, SW7 2AZ, UK}
\affiliation{School of Physics and Astronomy, Queen Mary University of London, London E1 4NS, UK}

\author{Justin C. Kasper}
\affiliation{BWX Technologies, Inc., Washington, DC 20001, USA}
\affiliation{Climate and Space Sciences and Engineering, University of Michigan, Ann Arbor, Michigan 48109, USA}

\author{SuPing Duan}
\affiliation{State Key Laboratory of Space Weather, Chinese Academy of Sciences, Beijing 100190, People’s Republic of China}

\begin{abstract}
	The Parker Solar Probe (PSP) provides us the unprecedentedly close approach observation to the Sun, and hence the possibility of directly understanding the "elementary process" which occurs
   in the kinetic scale of particles collective interactioin in solar coronal plasmas. We reported a kind of weak solar radio bursts (SRBs), which are detected by PSP when it passed a low-density
    magnetic channel during its second encounter phase. These weak SRBs have low starting frequecny $\sim 20$ MHz and narrow frequency range from a few tens MHz to a few hundres kHz. Their dynamic spectra display a strongly evolving feature of the intermediate relative drift rate decreasing
    rapidly from above 0.01/s to below 0.01/s. Analyses based on common empirical models of solar coronal plasmas indicate that these weak SRBs originate from the heliocentric distance $\sim 1.1-6.1~R_S$ (the solar radius), a typical
    solar wind acceleration region with a low-$\beta$ plasma, and indicate that their soruces have a typic motion velociy $\sim v_A$ (Alfv\'en velocity) obviously lower than that of fast electrons required by effectively exciting SRBs. We propose that solitary kinetic
    Alfv\'en waves with kinetic scales can be responsible for the generation of these small-scale
    weak SRBs, called solitary wave radiation (SWR).
\end{abstract} 


\section{Introduction} \label{sec-intro} 
Solar radio bursts (SRBs) are the most direct manifestation of energetic electrons, which exist ubiquitously in the solar atmosphere although their origin remains
poorly understood. It has been commonly believed that the ordinary type III and the type V SRBs are generated by energetic electrons via the process of magnetic reconnection during solar flare. The type II and the moving type IV SRBs are believed to be triggered by CME-associated shock accelerated electrons. The dynamic spectra of SRBs can provide the sensitive
and rich information of solar energetic electrons as well as of the background plasma in
the emitting source regions~\citep{Wild1950,Wildetal1954,Linetal1973,Melrose1980,Chenetal2017}.
For instance, a fast frequency drift of dynamic spectra of type III SRBs, characterized by a
relative frequency-drift rate $D\equiv \mid (df/dt)/f \mid>0.1/s$ ($f$ the emitting frequency), directly
indicates the travelling velocity of fast electron beams (FEBs) of emitting type III SRBs. The
dynamic spectra of type II SRBs usually present a slow drift with $D<0.01/s$ caused by
the energetic electrons accelerated by the CME driven shock. Moreover, the dynamic
spectra of moving type IV SRBs display often a very slow drift with $D\ll 0.01/s$ that can
be attributed to the sub-Alfv\'enic motion of coronal loops, in which energetic electrons
are trapped~\citep{Tanetal2019}.

For type III SRBs, using observations made by the Geotail and Akebono satellites, micro-type III bursts characterized by short lifetime, continuous, and weak emission are found by ~\citet{Moriokaetal2007,Moriokaetal2015}. They showed that micro-type III bursts have a distribution of emitted power flux that is different from that of ordinary type III bursts, and concluded that they are not just weaker versions of the ordinary bursts. 
Owing to PSP's close distance to the Sun, PSP/IS$\bigodot$IS observed a rich array of energetic particle events, which were not observed by the spacecraft at 1 AU, over the first two orbits~\citep{McComasetal2019}. \citet{Maetal2022} also found that, because of the radiation attenuation effect, many weak type III-like bursts with a higher cutoff frequency (hence narrower bandwidth) clearly detected by PSP can hardly be observed by WIND when PSP approaches its perihelion. In the microwave frequency range, the radio bursts called solar microwave drifting spikes, which are characterized typically by a short lifetime ($\sim$tens of ms), and an intermediate frequency drift rate ($\sim$a few hundred of MHz/s), have been detected by the Solar Broadband Radio Spectrometer of the National Astronomical Observatories of China. \citet{Wuetal2007} suggested that these Solar microwave drifting spikes probably are produced by accelerated electrons trapped within solitary kinetic Alfv\'en waves (SKAWs) potential well and the intermediate frequency drifts are attributed to the SKAWs propagation along the magnetic field.

Here, we report a kind of weak SRBs, which were found in recent observations of
the Parker Solar Probe (PSP) when it passed through a magnetic channel with low density at
a heliocentric distance of $\sim 36$ $R_S$ (the solar radius) during its second encounter
around the Sun~\citep{Foxetal2016,Pulupaetal2017,Pulupaetal2020,Maetal2021}.
These weak SRBs can be characterized by a weak intensity, low starting frequency, narrow frequency range, and
short lifetime, and probably originated from some small-scale emitting sources. Their dynamic spectra display a strongly evolving feature of that the relative drift rate decreases
rapidly from $D>0.01/s$ to $D< 0.01/s$ when the emitting frequency drifts downward from
a few tens MHz to a few hundreds kHz. Based on common empirical models of solar coronal
plasmas~\citep{MarianiandNeubauer1990,Leblancetal1998,WuandFang2003,WuandYang2007},
the observation of PSP shows that the emitting sources travel along the low-density magnetic
channel (or called equatorial coronal hole following ref.~\citet{Baleetal2019}) towards from
the heliocentric distance $R\sim 1.1~R_S$ to $R\sim 6.1~R_S$, a typical region of the solar
wind acceleration. 

The active region (AR) 12737, which is possibly associated with the low-density magnetic channel
observed by PSP, does not have evident flare or jet activities during the corresponding time period. In addition, there is no obvious Hard-X ray emission based on the observation of Fermi. Hence, it seems that the energetic electrons responsible for these weak SRBs not come directly from flare or jet activities in the AR 12737. However, in a low-$\beta$ plasma with $v_A>v_{T_e}$, where $v_A$ is the Alfv\'en velocity and $v_{T_e}$ is the electron thermal
velocity, SKAWs can travel at a velocity higher than $v_A$
and their electric fields may accelerate electrons to a velocity much higher than $v_A$ and trap
these energetic electrons within their potential wells. These trapped energetic electrons can
generate coherent radio radiation via the electron cyclotron maser (ECM) instability. In particular, the propagation and
evolution of SKAWs can reasonably explain the frequency drifting feature in the dynamic spectra
of these weak SRBs. Therefore, we propose that the kinetic-scale SKAWs can be responsible
for the generation of the weak small-scale SRBs observed by PSP, called solitary wave radiation
(SWR). The rest of the paper is organized as follows. In Section \ref{sec-ObsandAna}, the main observed properties of weak SRBs detected by PSP during its crossing of the low-density magnetic
channel and the correspongding solar wind plasma parameters are presented. Then, combining the in situ measurement of PSP and the empirical model of solar atmospheres, we foucus on discussing 
the generation mechanism of those weak SRBs in Section \ref{sec-ParaandMech}. Finally, summary and discussion are given in Section \ref{sec-SumandDis}.

\section{PSP Observation and Data Analysis} \label{sec-ObsandAna}
NASA's PSP~\citep{Foxetal2016} was launched into a heliocentric orbit on 2018 August 12
and will fly closer to the Sun than any other spacecraft before it. The primary science goals
for this mission are to trace the energy flow from the solar corona to the solar wind and to
help us understand the solar corona heating and the solar wind acceleration. In particular,
PSP is the first spacecraft to do in situ measurements of the solar corona and the source
region of the solar wind. To accomplish these science goals, four scientific instruments are
carried by PSP: Fields Experiment (FIELDS;~\citep{Baleetal2016}), Solar Wind Electrons
Alphas and Protons investigation (SWEAP;~\citep{Kasperetal2016}), Integrated Science Investigation
of the Sun (IS$\bigodot$IS;~\citep{McComasetal2016}), and Wide-field Imager for Solar PRobe
(WISPR;~\citep{Vourlidasetal2016}). Data presented in this study are mainly made by the
FIELDS experiment designed to do measurements of electric and magnetic fields. The radio
data are obtained by the Radio Frequency Spectrometer (RFS)~\citep{Pulupaetal2017} with
the dual-channel receiver, the Low Frequency Receiver (LFR): 10.5 kHz-1.7 MHz and the
High Frequency Receiver (HFR): 1.3 MHz-19.2 MHz.

\begin{figure*}[h!] 
    \begin{center}
    \includegraphics[scale=0.65,angle=0] {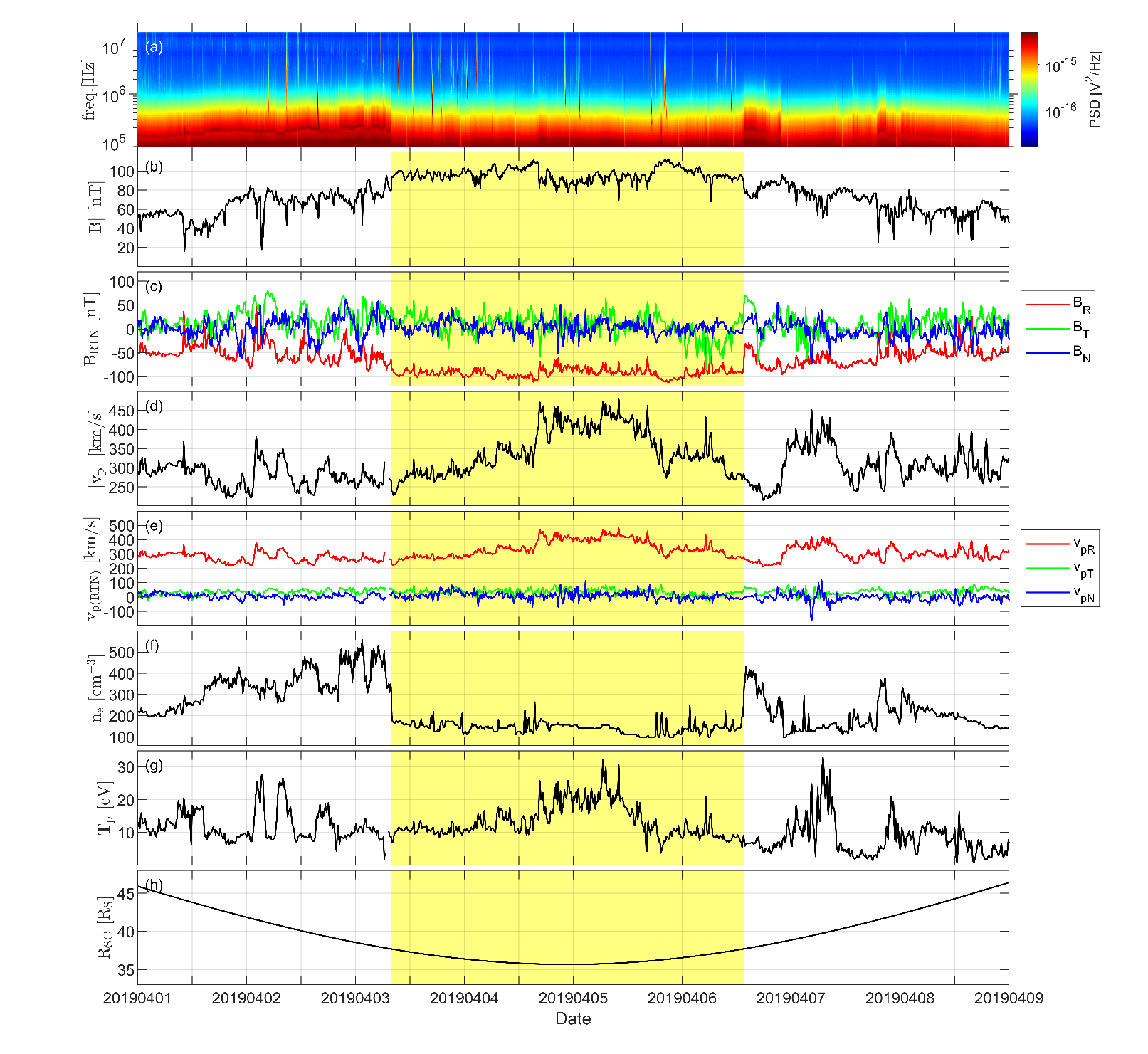}
    \caption{(color online) The solar wind plasma parameters observed by PSP between April 1 and
    9, 2019. From top to bottom, panels are the power spectral density (PSD)
    of the radio radiation (a), the magnetic field (b) and its components (c), the
    solar wind velocity (d) and its components (e), the plasma density (f) and
    temperature (g), and the heliocentric distance of PSP in units of the solar
    radius $R_S$ (h), respectively. A low-density magnetic channel is clearly
    displayed between 08:08:34 UT on April 3 and 13:17:09 UT on April 6,
    2019 (see the light yellow region).}
    \label{F1}
    \end{center}
\end{figure*}

The weak SRBs reported in this work were observed by PSP during its second solar encounter
(E02) from April 3 to 6, 2019, as shown in Figure \ref{F1}, in which panel (a) displays the power
spectral density (PSD) of the radio radiation, covering the frequency band from 10.5 kHz to
19.2 MHz. Panels (b) and (c) in Figure \ref{F1} present the magnetic field ($B$) and its three components
in the R-T-N frame ($B_R$, $B_T$, and $B_N$), and panels (d) and (e) do the solar wind velocity
($v_p$) and its components ($v_{pR}$, $v_{pT}$, and $v_{pN}$). Plasma density ($n_e$) and
temperature ($T_p$) are exhibited in panels (f) and (g), respectively, and the final panel (h)
shows the heliocentric distance of PSP in units of the solar radius $R_S$. From Figure \ref{F1}, one can
find that in the light yellow region between 08:08:34 UT on April 3 and 13:17:09 UT on April
6, 2019, the magnetic field $B$ strengthens mainly in its radial component $B_R$ from $\sim 80$
nT at the edge to $> 100$ nT in the center, and on the contrary the density $n_e$ significantly
reduces from $>400$ cm$^{-3}$ at the edge to $\sim 100$ cm$^{-3}$ in the center, as shown
in the panels (b-c) and in the panel (f), respectively. This implies that PSP was crossing a
low-density magnetic channel at the heliocentric distance $\sim 36~R_S$. On the other hand,
the panel (g) and the panels (d-e) show that the plasma temperature ($T_p$) and flow velocity
($v_p$, mainly its radial component $v_{pR}$), increase considerably from $\sim 10$ eV and
$\sim 250$ km/s at the edge to $\sim 30$ eV and $\sim 450$ km/s in the center, respectively,
implying that the plasma is heated and accelerated to form a solar wind stream in the center of
this open magnetic channel.

\begin{figure*}[h!] 
    \begin{center}
    \includegraphics[scale=0.6,angle=0] {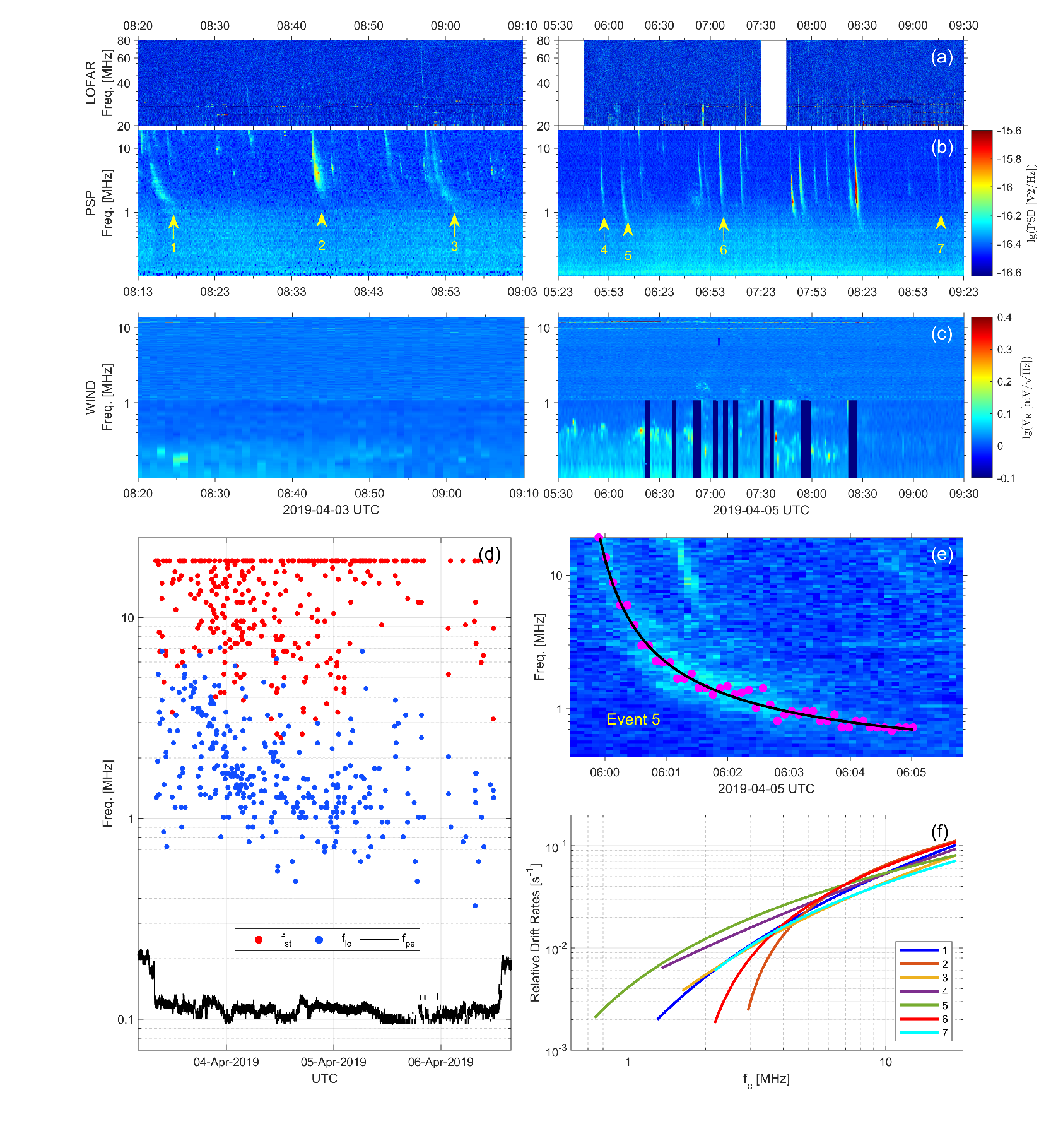}
    \caption{(color online) The weak SRBs observed by PSP during its crossing of the low-density magnetic
    channel. The top three rows show PSD measured by LOFAR (a), PSP (b), and
    WIND (c) in two intervals 08:13 to 09:03 on April 3, 2019 (left) and 05:23 to 09:23
    on April 5, 2019 (right). (d) The starting ($f_{st}$, red circles) and
    ending ($f_{lo}$, blue circles) frequencies for the 385 weak SRBs, and the local
    plasma frequency $f_{pe}$ (back line) along the PSP orbit. (e) The fitting curve (black curve) of maximal PSD for event 5 marked by arrow in the panel (b) through the least square fit, where the magenta dots are the location of maximal PSD in the dynamic spectrum. (f) The
    relative frequency-drift rate for the several typical bursts as marked by arrows
    in the panel (b).}
    \label{F2}
    \end{center}
    \end{figure*}

As seen in Figure \ref{F1}(a), a large number of SRBs presented in the observation of PSP during
its crossing of the low-density magnetic channel. We identified 428 SRBs with peak intensities
between $10^{-12.6}-10^{-16.3}$ V$^2$/Hz, above well the sensitivity of the RSF measurement,
$\sim 10^{-18}$ V$^2$/Hz~\citep{Pulupaetal2017}. Among these identifiable SRBs only 43
bursts ($\sim 10\%$) have peak intensities higher than $10^{-15}$ V$^2$/Hz, and all the
rest 385 bursts have peak intensities lower than $10^{-15}$ V$^2$/Hz. These very weak SRBs have a
rather high occurrence rate of $\sim$ five bursts per hour in average and they can be regarded as point sources and should originate
from small-scale emitting sources because of their weakness. In the
panel (d) of Figure \ref{F2}, the starting ($f_{st}$) and ending ($f_{lo}$) frequencies of these weak SRBs are displayed by red and blue circle dots, respectively. The local plasma frequency $f_{pe}$ along the PSP orbit is presented
by the black line. One can see that the lowest ending frequency ($f_{lo}=367$ kHz) of these
weak SRBs is well higher than the local plasma frequency $f_{pe}\sim 100$ kHz, implying
that they all had stopped radiating before arriving at PSP. The starting frequencies of 247
weak SRBs ($\sim 64\%$) can be determined definitely because their $f_{st}$ all are well
lower than the upper-limit frequency of the RFS measurement (i.e., $\sim 19.2$ MHz). Other
about one third (138 events) weak SRBs, however, have $f_{st}$ reaching the upper-limit frequency of the RFS measurement
as shown in Figure \ref{F2}(d), which indicates that their actual starting frequencies possibly are higher
than $19.2$ MHz.

In order to find their real starting frequencies, we further compared the PSP observation
with that by LOFAR (the Low-Frequency Array in Europe) during the same interval, which
has an effective-observation frequency range between 20--80 MHz. However, due to limited observation time of LOFAR allocated for solar and spaceweather division, only 151 of these 385 weak
SRBs occurred in the solar observation windows of LOFAR. The result showed
that only less than one fourth (36 of the 151 SRBs) have $f_{st}>20$ MHz and hence were
observed by LOFAR. Figure \ref{F2} shows further the comparison between observations of LOFAR
(a), PSP (b), and WIND (c) for two intervals 08:13 to 09:03 on April 3, 2019 (left) and 05:23
to 09:23 on April 5, 2019 (right), in which there are 5 (left) and 16 (right) weak SRBs, respectively.
In the panels (a) and (c) of Figure \ref{F2} an 7-minutes delay for LOFAR and WIND is due to the consideration
of the radiation propagation from PSP to LOFAR and WIND. In the comparison with the
LOFAR observation, only part bursts extend their emission to frequencies higher than 20
MHz as seen in Figure \ref{F2}(a). In the WIND observation, however, almost all these weak SRBs are
invisible because of their weakness as seen in Figure \ref{F2}(c). In fact, only some strong SRBs with
the peak intensity higher than $10^{-15}$ V$^2$/Hz can be seen by WIND, and these strong
SRBs, in general, also have a higher starting frequency in the LOFAR observation.

The relative frequency-drift rate is an important and critical parameter that reflects directly
the environment and motion of the emitting source in the background plasma. It has been
used as a typically characteristic parameter to classify various kinds of SRBs. Here, we take the event 5 marked by arrow in Figure \ref{F2}(b) as an example to show the determination of the relative frequency-drift rate $D$. After subtracting the background noise, the dynamic spectrum of the event 5 is shown in Figure \ref{F2}(e), in which at the low frequencies ($\sim$1 MHz or below) the burst becomes more clear compared to the background. By finding the frequencies of maximal PSD (magenta dots) of the event 5 at each time point, we can obtain the fitting curve (black curve) for the frequency-time relation in a polynomial form, $\lg f=at^b+c$, with the parameters $a=5.722$, $b=-0.704$, and $c=5.474$ for the event 5, which are obtained by the least square fit. Then, the relative frequency-drift rate can be calculated by the expression of the fitting curve. The advantage of this method is to avoid limiting by lower time resolution around the highest frequencies ($\sim$19 MHz) in the determination of drift rate. Figure \ref{F2}(f) shows
the relative frequency-drift rate $D$ versus the central frequency for the several typical examples
marked by arrows in Figure \ref{F2}(b), and the result shows that their relative drifting rates have an
initial value lower than and close to $\sim 0.1/s$ and rapidly slow down from $>0.01/s$ to
$<0.01/s$ as the emitting frequency drifts downward from a few tens MHz to a few hundreds
kHz. This implies that their emitting sources probably experienced strong dynamical evolutions.
An interesting and puzzling question is where the energetic electrons responsible for these
weak SRBs come from, which should exist rather ubiquitously.

On the other hand, combining the remote sensing observations by the Hinode EUV Imaging
Spectrometer and the Solar Dynamics Observatory Atmospheric Imaging Assembly, \citet{Harraetal2021} found that the active region (AR) 12737 is possibly associated
with the low-density magnetic channel observed by PSP. The AR 12737, however, does not
have evident flare or jet activities during the corresponding time period. Based on the observation of Fermi, there is also no obvious Hard-X ray emission. Therefore, it seems
to be unlikely that FEBs responsible for these weak SRBs come directly from flare or jet
activities in the AR 12737. In addition, there is an extended blue-shifted outflow region inside
the AR 12737, and its expanding behavior probably leads to the formation of the low-density
magnetic channel and the solar wind stream. Moreover, the solar wind magnetic field is
dominated by the radial component $B_R$ and follows a similar Parker spiral structure~\citep{Parker1958}.
Therefore, it is a reasonable inference that the magnetic channel extends from the AR 12737
in the solar corona to the solar wind. Accompanied with the acceleration and heating of the
solar wind, the emitting sources of these weak SRBs were formed and travelled outwards
along this magnetic channel~\citep{ReinerandKaiser1999,Maetal2021}. In particular, the
rapid decrease of their relative drifting rates indicates that their emitting sources should
have experienced strong dynamic evolutions in the solar wind acceleration region. 

\section{Plasma Parameters Model and Generation Mechanism of Weak SRBs} \label{sec-ParaandMech}
\subsection{Plasma Parameters Model} \label{subsec-Parameters}
Both the emitting source and the emission mechanism of these weak SRBs sensitively
depend on the local plasma parameters in the source region. For example, their emitting
frequencies are associated closely to be the characteristic frequencies of the local plasma
in the source regions, such as the plasma frequency $f_{pe}=8.98\sqrt{n_e}$ kHz for the
plasma emission and the electron cyclotron frequency $f_{ce}=2.8B_0$ MHz for the ECM
emission, where the ambient plasma density $n_e$ and magnetic field $B_0$ are in units
of cm$^{-3}$ and Gauss, respectively. The relative frequency-drift rate $D$ of the dynamic
spectra, in general, can be determined by the moving velocity of the emitting source and
the variation of the characteristic frequencies ($f_{pe}$ or $f_{ce}$), which are given directly
by the ambient plasma density ($n_e$) or magnetic field ($B_0$) along the propagation
path of the weak SRBs.

A widely adopted model for the radial distribution of the average plasma density from the
solar corona at $\sim 1.8~R_S$ to the solar wind at $\sim 1$ AU is the polynomial distribution
proposed by \citet{Leblancetal1998}, $n_e=a\left(242.5r^{-6}+12.5r^{-4}+r^{-2}\right)10^5$
cm$^{-3}$, in which $r$ is the heliocentric distance in units of the solar radius $R_S$, the first
and third terms proportional to $r^{-6}$ and $r^{-2}$ are dominant in the solar corona and
the solar wind, respectively, the second term proportional to $r^{-4}$ is used to fit the transition
between the corona and the wind, and the coefficient $a$ can be determined by the measured
density value at 1 AU or other distances. In the low corona, however, the density gradient is
very steep in the exponential fall way. In order to fit the density of $\sim 10^{10}$ cm$^{-3}$
at the base of the corona, \citet{WuandFang2003} introduced an exponential function
with a scale height $h\sim$ 0.02 $R_S$, $10^{10}e^{-50(r-1)}$ cm$^{-3}$, to model the density
distribution in the low corona. In addition, a density dilution factor, $d_f(r)\equiv \left[1+9e^{-\left(r-1\right)^2/100}\right]^{-1}$,
may be invoked to describe the low-density feature of the magnetic channel~\citep{EsserandSasselov1999,Esseretal1999,Youngetal1999,Teriacaetal2003,WuandYang2007}.
In consequence, the radial distribution of the electron density along the magnetic channel can
be fitted by
\begin{eqnarray}
n_e(r)=d_f(r)\left[10^{10}e^{-50(r-1)}+(388r^{-4}+20r^{-2}+1.6)10^5r^{-2}\right](\mathrm{cm^{-3}}).  \label{n}
\end{eqnarray}
On the other hand, some numerical two-fluid models of high speed streams from the corona
to the solar wind~\citep{Huetal1997} showed that the radial distribution of the electron temperature
can be characterized by a quick increase from $\sim 5\times 10^5$ K to $\sim 1.5\times 10^6$
K within $r<3$, then a slow decrease by a factor of $\sim 10$ at a radial distance of about a
few tens of $R_S$, followed by a slower decrease in the interplanetary space. Such temperature
behavior may be described characteristically by
\begin{equation}
 T_e(r)=\left[4.2re^{1-r/3}+6(3r^{-1})^{0.3}\right]10^5~(\mathrm{K}). \label{T}
\end{equation}
The radial distribution of the magnetic field along the magnetic channel can be modelled
by the combination of a dipole field $\propto r^{-3}$ in the corona and a monopole field
$\propto r^{-2}$ in the solar wind as \citep{MarianiandNeubauer1990},
\begin{equation}
B_0(r)=15r^{-3}+r^{-2}~(\mathrm{G}). \label{B}
\end{equation}
Here, the averaged electron density ($n_e\sim 120~\mathrm{cm^{-3}}$), temperature
($T_e\sim 25~\mathrm{eV}$), and magnetic field ($B_0\sim 100~\mathrm{nT}$) in situ
measured by PSP at $\sim 36.6~R_S$ \citep{Halekasetal2020} have been used to fit the
coefficients in the equations (\ref{n})--(\ref{B}).

\begin{figure*}[h!]
    \begin{center}
    \includegraphics[scale=0.6,angle=0] {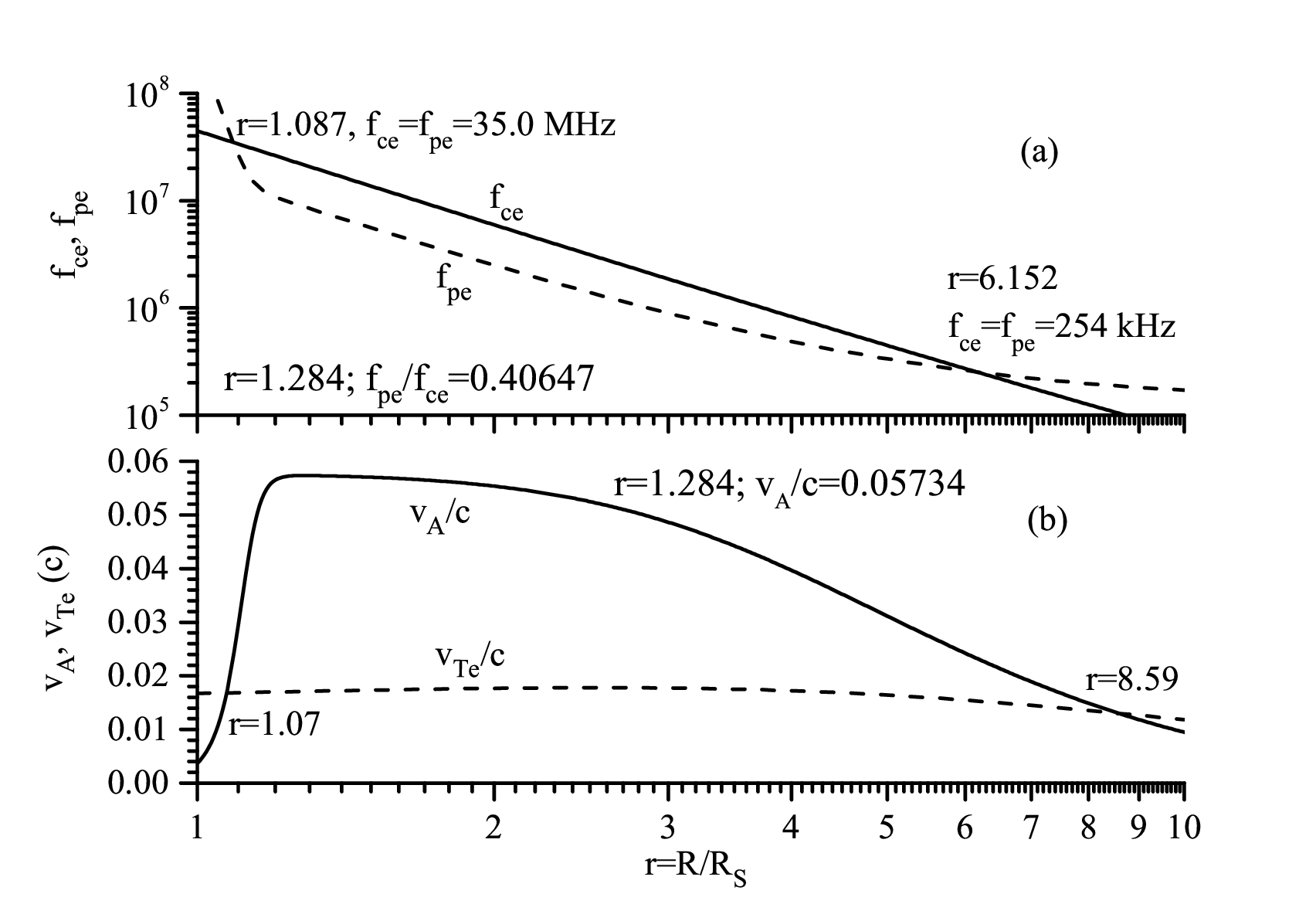}
    \caption{The radial distributions of characteristic frequencies and velocites. (a) The radial distributions of the electron cyclotron ($f_{ce}$) and plasma
                    ($f_{pe}$) frequencies. (b) The radial distributions of the Alfv\'en ($v_A$) and electron thermal
                    ($v_{T_e}$) velocities.}
    \label{F3}
    \end{center}
    \end{figure*}

Based on the empirical models mentioned above, Figure \ref{F3}(a) shows the radial distributions
of the electron cyclotron ($f_{ce}=2.8\times10^6 B_0(r)~\mathrm{Hz}$; the solid line) and
plasma ($f_{pe}=8.98\times10^3 \sqrt{n_e(r)}~\mathrm{Hz}$; the dashed line) frequencies,
respectively. From Figure \ref{F3}(a), it can be found that the low-density magnetic channel extends
from $\sim1.1$ to $\sim 6.1~R_S$ in which the electron cyclotron frequency is evidently
higher than the plasma frequency, that is, $f_{ce}>f_{pe}$ and the corresponding range of
the characteristic frequencies between $\sim 30$ MHz and $\sim 300$ kHz covers well the
emitting frequency band of the weak SRBs. This further confirms that their emitting sources
can be well located within the extended solar corona from $\sim 1.1$ to $\sim 6.1~R_S$,
which is a low-density magnetic channel or an equatorial coronal hole with outflows and
open magnetic fields. Usually, this extended coronal region is extensively believed to be the
important region of the solar wind origin and acceleration.

Figure \ref{F3}(b) presents the local Alfv\'en velocity $v_A$ (solid line) and the electron thermal
velocity $v_{T_e}$ (dashed line) normalized by the light velocity $c$. From Figure \ref{F3}(b) one
has $v_A>v_{T_e}$ in the emitting source region of these weak SRBs ($\sim1.1-6.1~R_S$),
implying there is a low-$\beta$ plasma of $\beta<m_e/m_p$ in the source region, where $m_e$ and $m_p$ are electron and proton masses, respectively. Based on
the theory of kinetic Alfv\'en waves (KAWs)~\citep{Wu2012,WuandChen2020}, a low-$\beta$ SKAW, accompanied by a dip
density soliton with an inner density $n_{em}<n_0$, can propagate at a super-Alfv\'en
velocity $v_{sw}=\sqrt{\left(2+n_m\right)/3n_m}v_A>v_A$ along the magnetic field, where
$n_m\equiv n_{em}/n_0$ is the inner density normalized by the ambient density $n_0$. In
particular, the field-aligned electric field $E_\parallel$ of SKAW can efficiently trap and accelerate
electrons to a typical velocity $v_{ez}=\left(n_m^{-1}-1\right)v_{sw}$, much higher than
$v_A$ for $n_m\ll 1$~\citep{Wu2012,WuandChen2020}. For example, for $n_m=0.2$, one has
$v_{sw}=\sqrt{11/3}v_A\simeq 1.915v_A$ and $v_{ez}=4v_{sw}\simeq 7.66v_A$. This
can provide an efficient acceleration mechanism for the local generation of energetic electrons
that is required to excit the emissions of the weak SRBs.

\subsection{Emission mechanism of weak SRBs} \label{subsec-mechanism}
It is worth noting that the low-density magnetic channel has similar plasma conditions to
the source region of the terrestrial auroral kilometric radiation (AKR, see e.g. refs.~\citet{Mozeretal1980,Bryant1990,WuandChao2004}),
in which the electron cyclotron frequency $f_{ce}$ is higher than the plasma frequency $f_{pe}$
and the Alfv\'en velocity $v_A$ is larger than the electron thermal velocity $v_{T_e}$. The
condition $f_{ce}>f_{pe}$ indicates that the coherent radio radiation at the cyclotron frequency
and its harmonics can be effectively excited by the ECM instability.

The solar wind acceleration region, located in the extended solar corona at $\sim 1.1-10$
$R_S$, is a complex dynamical transition region, where the coronal plasma is heated and
accelerated into the solar wind. Alfv{\'e}n wave (AW) turbulence, originating from the
photospheric turbulence and convection, plays an important role in the coronal heating
and solar wind acceleration~\citep{CranmerandBallegooijen2005,Cranmeretal2007}. The
AW turbulence, via the anisotropic turbulent cascade, period doubling and wave breaking~\citep{GoldreichandSridhar1995,Tsurutanietal2018},
can extend into the kinetic scales of particles and become the kinetic AWs (i.e., KAWs). Meanwhile, in a low-$\beta$ plasma of
$v_A>v_{T_e}$, nonlinear solitary wavelets of KAWs (i.e., SKAWs), may be formed
effectively because KAWs can be free from the heavy Landau damping when their phase
speed ($>v_A$) is considerably larger than the electron thermal speed ($v_{T_e}$).
SKAWs and their associated structures have been observed and identified extensively
in near-earth space plasmas, such as in the auroral plasma. A number of studies both
in observation and theory have shown that the nonlinear KAWs can play an important
role in the field-aligned acceleration of electrons and the crossing-field heating of ions in
the aurora plasma~\citep{Louarnetal1994,Wahlundetal1994,Wuetal1995,Chastonetal1999,Stasiewiczetal2000,WuandChao2004}.
For the case of the solar corona, also it is found that SKAWs can be responsible for the
anomalous anisotropic energization of minor heavy ions discovered by the Solar and
Heliospheric Observatory (SOHO) in the extended solar corona of $\sim1.5-5~R_S$~\citep{WuandYang2007,Wu2012,WuandChen2020}.

On the other hand, the field-aligned electric field of SKAWs has a typical dipole structure,
which can accelerate electrons along the magnetic field and trap these energetic electrons
inside the potential well. Some recent works have shown that the presence of AW turbulence
may significantly influence the ECM instability~\citep{Wuetal2012,Wu2014,Zhaoetal2015,Chenetal2017,Chenetal2021}.
Recently \citet{Kasperetal2021} found that the strong AW turbulence exists
not only in super-Alfv\'enic streams such as in the solar wind but also can present in the
low-$\beta$ solar corona with a sub-Alfv\'enic stream. They reported the spectrum of
Alfv\'en turbulence measured by PSP during its eighth solar encounter (E08), which has
a typical power spectral density from $\sim 10^6$ to $10^3$ km$^2$/s$^2$/Hz and a
spectral index $\sim -3/2$ for the inertial turbulence range from 0.002 to 0.2 Hz. An ambient
Alfv\'en velocity $v_A\sim 450$ km/s indicates the relative energy density of the Alfv\'en
turbulence $\delta_B\sim 2\times 10^{-2}$~\citep{Kasperetal2021}. However, this relative
strength expresses only an average turbulence level of AWs, while the actual relative strength
can vary dependently on cases in a wide range, say, 0.01 to 0.1, in the solar corona~\citep{CranmerandBallegooijen2005}.

\begin{figure*}[h!]
    \begin{center}
    \includegraphics[scale=0.6,angle=0] {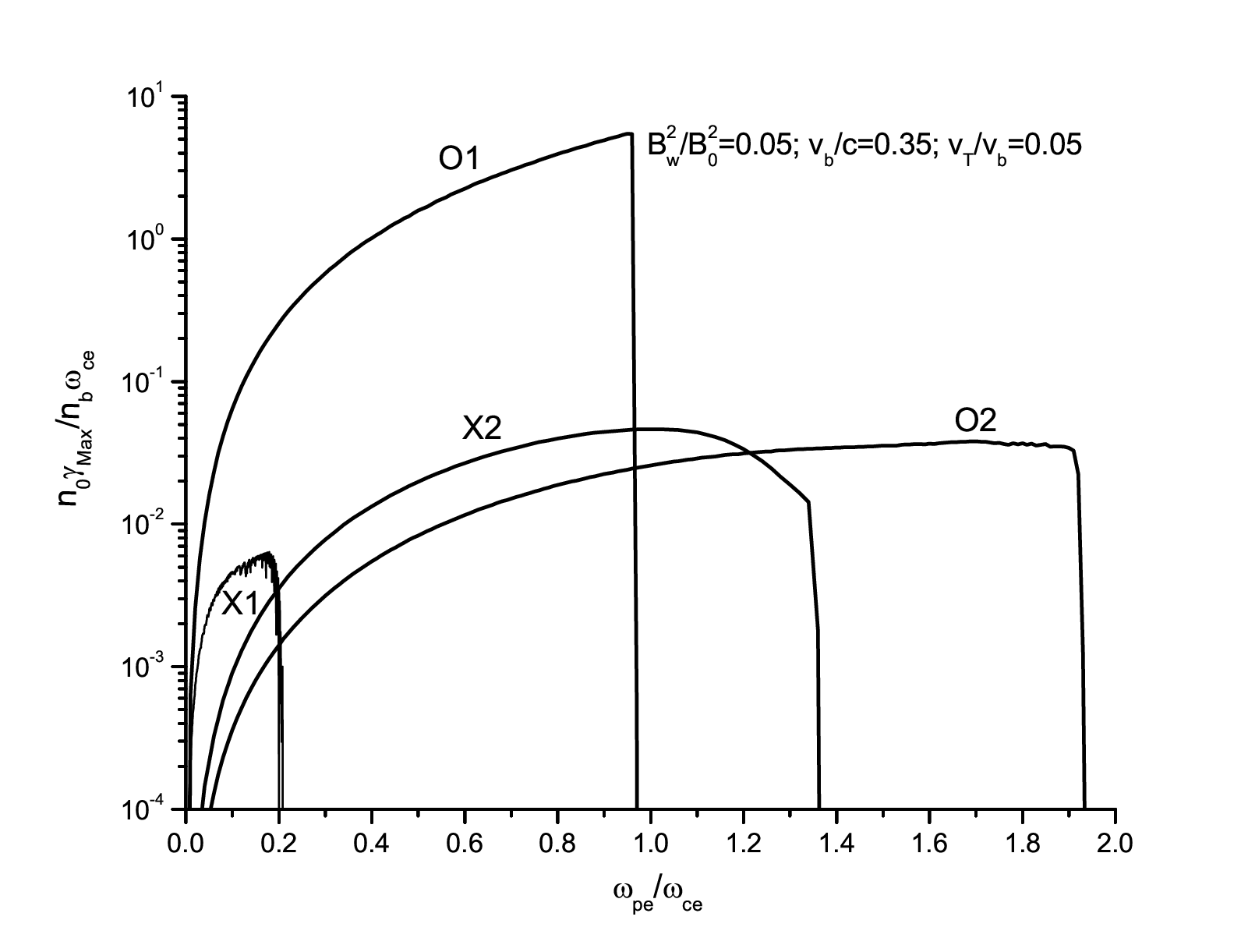}
    \caption{The growth rates of the ECM instability excited by the crescent-shaped
    distributions in Equation (\ref{F}). O1 (X1) and O2 (X2) denote the ordinary
    (extraordinary) mode in the fundamental and harmonic frequencies, respectively.
    Parameters $\delta_B=0.05$, $v_b/c=0.35$, and $v_T/v_b=0.05$ have been used.}
    \label{F4}
    \end{center}
    \end{figure*} 

Following \citet{Wuetal2012}, the velocity distribution function of beam electrons
with a characteristic beam velocity $v_b$ under the influence of AW turbulence with the
relative strength $\delta_B$ can be modelled by the so-called crescent-shaped distribution
\begin{equation}
f_b(v,\mu)=A\exp\left[-{\left(v-v_b\right)^2\over v_T^2}-{1-\mu^2\over \theta_\perp^2}\right], \label{F}
\end{equation}
where $A$ is the normalization constant, $\mu\equiv v_z/v$ is the pitch angle of the electron
velocity, $v_T$ is the velocity spread of the beam electrons, and $\theta_\perp\equiv\sqrt{v_T^2/v_b^2+2\delta_B}$
is the pitch-angle spread of the beam electrons. In particular, the crescent-shaped distribution
can effectively excite the ECM emission~\citep{Wuetal2012,Wu2014,Zhaoetal2015}. Figure \ref{F4}
shows the growth rates of the ECM instability versus the frequency ratio $\omega_{pe}/\omega_{ce}$,
where the parameters $v_b=0.35 c$, $v_T=0.05v_b$, and $\delta_B=0.05$ have been used.
As shown in Figure \ref{F4}, the excited modes depend considerably on the frequency ratio $\omega_{pe}/\omega_{ce}$,
in which the most easily excited emission is the fundamental ordinary mode (O1) in the
low-$\beta$ plasma of $\omega_{pe}/\omega_{ce}<1$, while for the case of $1<\omega_{pe}/\omega_{ce}<2$,
the harmonic waves of the ordinary (O2) and extraordinary (X2) modes also may be excited
but at a much lower growth rate. However, the fundamental extraordinary mode (X1) can
be excited only in the extreme condition of $\omega_{pe}/\omega_{ce}\ll 1$ because of its
higher cutoff frequency than that of the ordinary mode.

As shown in Figure \ref{F3}(a), in the low-density magnetic channel between $\sim1.1$ and $\sim 6.1~R_S$,
the exciting condition for the ECM emission, $\omega_{pe}/\omega_{ce}<1$, can be satisfied
and the frequency range between $\sim 35$ MHz and $\sim 0.25$ MHz covers well the
emitting frequencies of the weak SRBs. Meanwhile, the low-$\beta$ condition of $v_A>v_{T_e}$
in the emitting source region indicates that SKAWs with a dip density soliton can propagate
at a super-Alfv\'en velocity $v_{sw}>v_A$ along the magnetic field~\citep{Wu2012,WuandChen2020}.
In particular, their field-aligned electric field $E_\parallel$ can efficiently accelerate electrons
to form an oscillating energetic electron beam with a higher velocity $v_{ez}>v_{sw}$. For
SKAW with a normalized inner density $n_m<1$, the energetic electron beam can have a
characterized beam velocity $v_b=v_{ez}$~\citep{Wu2012,WuandChen2020}, that is,
\begin{equation}
v_b=v_{ez}={1-n_m\over n_m}v_{sw}={1-n_m\over n_m}\sqrt{2+n_m\over 3n_m}v_A. \label{Vb}
\end{equation}
We propose that the ECM emission excited by the energetic electron beam, trapped in the
potential well of the SKAW, can be responsible for the weak SRBs, called SWR, and while
the frequency drift of SWR is caused by the travel of the SKAW at the velocity $v_{sw}=\sqrt{\left(2+n_m\right)/3n_m}v_A>v_A$. Specificly, a single small-scale weak SRB is attributed to a single SKAW. 

\begin{figure*}[h!]
    \begin{center}
    \includegraphics[scale=1.2,angle=0] {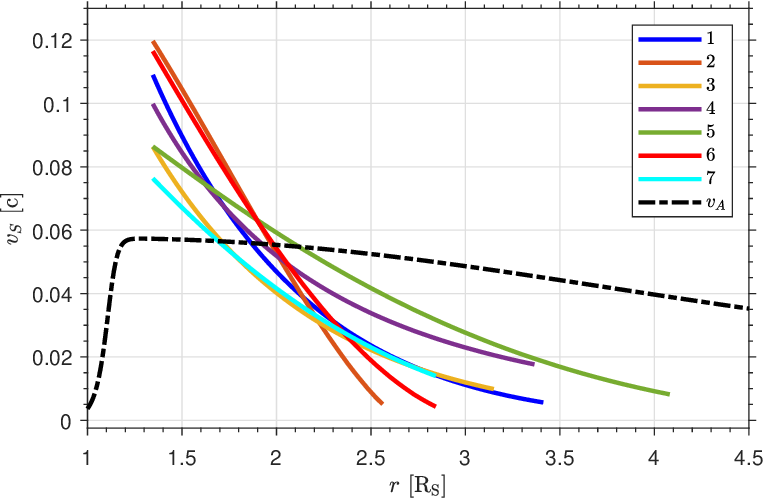}
    \caption{(color online) The radial distributions of the emitting source velocities for the typical
    events of weak SRBs $v_S/c$ (solid lines) and the Alfv{\'e}n velocity $v_A/c$
    (dashed line) along an open magnetic field region in the low-density magnetic
    channel, where all velocities have been normalized by the light speed $c$.}
    \label{F5}
    \end{center}
    \end{figure*}

Combining the relative frequency-drifting rate $D$, presented in Figure \ref{F2}(f), with the radial
distribution of the electron cyclotron frequency $f_{ce}$ (i.e., $D=(df_{ce}/dr)(v_S/f$)), Figure \ref{F5} shows the moving
velocity of the emitting sources $v_S/c$ versus the heliocentric distance (solid lines) for the
several typical weak SRBs in Figure \ref{F2}(b), where the local Alfv{\'e}n velocity $v_A/c$ is presented
by the dashed line for the sake of comparison, with a typical value $v_A\sim 0.05c$ in the
source region. The sources have initial velocities $v_S\sim 0.1 c\sim 2v_A$ and then rapidly
descend to $v_S<v_A$. Here, we point out that these weak SRBs were excited by the energetic electrons with velocity $v_b\sim0.35 c\sim 7 v_A$ which are trapped in the potential well of SKAW with velocity $v_{sw}=v_S\sim 0.1 c\sim 2v_A$ (for $n_m=0.2$). From Figure \ref{F5} it can be found that the motion of these sources has a common
evolutionary feature, that is, a rapid deceleration with the heliocentric distance from initial
$v_S>v_A$ in the closer to the Sun to $v_S<v_A$ in further distances, implying that the emitting
sources experience an evident and lasting deceleration as their traveling outwards. This common
feature can be reasonably explained by the evolving property of SKAWs due to the dissipation
in low-$\beta$ plasma~\citep{Wahlundetal1994,VoitenkoandGoossens2000,Wuetal2007}.

In the initial stage of the SKAW formation, the energetic electrons with the characteristic
oscillating velocity $v_b$, accelerated by the SKAW electric field, are trapped within the
SKAW potential well and travel at the velocity $v_{sw}$ together with the SKAW~\citep{Wuetal2007}.
It is these trapped energetic electrons that trigger SWR via the ECM mechanism, in which
the transverse free energy required by exciting the ECM instability can be provided by the
wave-particle scattering of AW turbulence. As propagating, the SKAW may loss energy due
to various possible dissipations, such as the Coulomb collision or ion-acoustic turbulence~\citep{Wahlundetal1994,VoitenkoandGoossens2000,Wuetal2007}.
This leads to the deceleration of the SKAW, i.e., the emitting source, because the propagating
velocity $v_{sw}$ decreases as the dissipation of the SKAW. In the meantime, accompanying
with the dissipation, the symmetry of the SKAW potential well is deformed and evolves into
a shock-like structure, and in consequence the part energetic electrons may escape from the
SKAW potential well, until the SKAW is exhausted when the inner density $n_m\to 1$~\citep{Wahlundetal1994,WuandChao2004}.
The energetic electrons escaping from the dissipated SKAW gradually merge into the local
plasma environment with strong AW turbulence and the corresponding SWR has a wider
spectral band and a slower drifting velocity, and ultimately travels together with the accelerated
solar wind at a velocity much lower than the local Alfv\'en velocity as shown in Figure \ref{F5}.

The dissipation time of SKAWs may be estimated approximately by the inverse of the damping
rate $\gamma\sim 0.5\nu_e k_\perp^2\lambda_e^2$ in the coronal plasma~\citep{VoitenkoandGoossens2000,Wuetal2007},
where $\nu_e=2.91\times10^{-6}n_e T_e^{-1.5}\ln\Lambda$ is the electron collision frequency
and $\ln\Lambda$ is the Coulomb logarithm. Assuming the typical electron density $n_e\sim10^5$
cm$^{-3}$ and temperature $T_e\sim 150$ eV in the source region, one has $\ln\Lambda\sim 20$
and $\nu_e\sim 3.16\times 10^{-3}$ Hz. In a low-$\beta$ plasma of $v_A>v_{T_e}$, SKAWs have
a characteristic width $\lambda_\perp\sim 2\pi\lambda_e$ (i.e., $k_\perp\lambda_e\sim 1$),
where $\lambda_e\equiv c/\omega_{pe}$ is the electron inertial length~\citep{Chastonetal1999}.
In consequence, the damping rate is $\gamma\sim 1.58\times10^{-3}$ Hz, implying that the
dissipation time of SKAWs typically is about ten minutes (i.e., $\tau\sim\gamma^{-1}\sim 600$ s).
However, actual damping rate in general can be higher than this collision damping, and hence
the observed lifetime of SWR may be shorter than the estimation here. A single SKAW can contribute to a single burst and the energetic electrons trapped in the potential well of SKAW can gain energy by the acceleration of the SKAW electric field. The power of the SKAW may be estimated by the production of the Joule heating rate ($\mathbf{J} \cdot \mathbf{E}$) in the SKAW and the SKAW volume, that is, $Q_H\sim jE\lambda_\perp^2\lambda_\parallel\sim2.9\times10^{11}$ W, where $j\sim en_{em}v_b$ and $E\sim E_m\sqrt{m_e/m_p}v_AB_0$ are the current density and the electric field in the SKAW. Here, the parallel scale of SKAW $\lambda_\parallel\sim10^3\lambda_\perp$, the dimensionless parameters $n_m=0.2$ (hence, $v_b\simeq 7.66v_A\sim0.35c$) and $E_m=3.5$ ~\citep{Wu2012,WuandChen2020} have been assumed. The magnetic field $B_0\simeq1.12$ G is obtained from Eq. (3) for the normalized heliocentric distance $r=2.5$ has been used. On the other hand, the rate of energy loss due to the radio emission may be estimated as $Q_R\sim P\vartriangle fR_{pb}^2/(Z_0L_{eff}^2\Gamma^2)\sim 1.4\times10^9$ W $\ll Q_H$ (i.e., $Q_R\sim0.01~Q_H$), where the effective antenna length $L_{eff}\sim 1$ m, capacitive gain factor $\Gamma=0.32$, and impedance of free space $Z_0=377~\rm{\Omega}$ have been used ~\citep{Pulupaetal2017,Jebarajetal2023}. Here, taking event 5 for example, we have adopted the averge power spectral density $P\sim10^{-16}$ V$^2$/Hz, the frequency width $\vartriangle f\sim10^6$ Hz, and the distance between PSP and emitting source $R_{pb}\sim 33.5~R_S$. If we consider further the fact that the coherent radio radiation often is a strong anisotropic emission, the radiation will be only concentrated within a small flare angle which cross section is possible much smaller than $R_{pb}^2$. In consequence, $Q_R$ will decrease further and become much smaller than $Q_H$. In fact, the energy loss via the radio radiation is only a very small part of the energy loss of the SKAW, and the major energy loss of the SKAW is due to the dissipation caused by the classical or abnormal collision.

\section{Summary and Discussion} \label{sec-SumandDis}
In summary, we reported a kind of weak SRBs observed by PSP when crossing
a low-density magnetic channel during its E02 phase. These weak SRBs have a weak intensity
lower than $10^{-15}$ V$^2$/Hz, the relatively low starting frequency ($\sim 20$ MHz) and a narrow frequency range from a few tens MHz to a few hundreds kHz, and an evolving and intermediate relative frequency-draft rate $D\equiv \mid (df/dt)/f \mid$ from $D>0.01$/s to $<0.01$/s. They can occur quite frequently (five bursts per hour) and 
the nature of their weak intensity indicates that they originate from small-scale emitting sources.
Based on the common empirical models for the solar coronal plasma, these small-scale emitting
sources lied in the heliocentric distance between $\sim 1.1-6.1~R_S$, a typical solar wind
acceleration region. We proposed that SKAWs in kinetic scales, which are formed easily
in the solar wind acceleration region with a low-$\beta$ plasma environment, can be responsible
for the small-scale emitting sources of these weak SRBs, called SWR (solitary wave radiation).

Although the radio radiation of SWR has insignificant impact on the space plasma environment,
the kinetic-scale characteristic of their emitting sources has important implications on the
dynamics of magnetic plasmas in the solar wind acceleration region. One of the unsolved
problems in solar physics is the heating and acceleration mechanism of coronal plasmas into
the solar wind in the extended corona from 1.1 to 10 $R_S$. The complexity of the extended
coronal plasma both in the kinetics and dynamics is the result of the plasma density decreasing
with the heliocentric distance as well as the complicated magnetic topology in coronal plasmas.
The decrease of the density leads to the transition of the coronal plasma from a collisionally
dominated plasma to nearly collisionless one. As a result, the kinetic wave-particle interaction
process plays an important role in the heating and acceleration of the coronal plasmas. On
the other hand, the fully ionized state of the hydrogen, the major solar atmospheric component,
results in the impossibility to gain the physical information associated with the acceleration
and heating processes via spectral line observations of the hydrogen, which is a main method
of inferring the physical situation and processes in the potospere and chromosphere. Alternately, however,
radio observations, especially the observations of radiation originating from small-scale
emitting sources can provide us rich information of energetic electrons and their kinetic
processes as well as of the ambient magnetic plasmas in the solar wind acceleration region.

The present research at PMO was supported by the Strategic Priority Research Program of the Chinese Academy of Sciences under grant No. XDB0560000, the National Natural Science Foundation of China (NSFC) under grant Nos. 42174195, 11873018 and 11790302, and supported by the Specialized Research Fund for State Key Laboratories. We acknowledge the NASA Parker Solar Probe Mission and the FIELDS team led by S. D. Bale and the SWEAP team led by J. C. Kasper for use of data; the International LOFAR Telescope team for usage of the data. The FIELDS and SWEAP experiments on the Parker Solar Probe spacecraft were designed and developed under NASA contract NNN06AA01C. The data could be obtained on the website: https://spdf.gsfc.nasa.gov/pub/data/psp/. We thank the anonymous referee for the useful comments which led to an improved version of the manuscript.


\end{document}